\documentclass[10pt]{article}

\usepackage[T1]{fontenc}
\usepackage[utf8]{inputenc}
\usepackage[english]{babel}
\usepackage{amsmath,latexsym,amsfonts,amsthm,bm,mathtools}
\usepackage{graphicx}
\usepackage[top=1.5in, bottom=1.5in, left=1.25in, right=1.25in]{geometry}
\usepackage{booktabs,url,tikz}
\usepackage[makeroom]{cancel}
\usepackage{dsfont}
\usepackage{relsize}
\usepackage{multirow}

\usepackage{algorithm}
\usepackage{algpseudocodex}
\usepackage{bbm}
\usepackage{subcaption}

\newcommand{\Exp}[1]{\mathbb{E}\left[ #1\right]}
\newcommand{\Var}[1]{\mathbb{V}\textrm{ar}\left[ #1\right]}
\newcommand{\approxx}{\overset{\sim}{=}}

\newcommand{\indep}{\raisebox{0.05em}{\rotatebox[origin=c]{90}{$\models$}}}

\title{Bayesianize Fuzziness in the \\Statistical Analysis of Fuzzy Data}
\author{Antonio Calcagn\`{i}$^{1\ast}$, Przemys{\l}aw Grzegorzewski$^{2}$, Maciej Romaniuk$^{3}$ \\\\
		\footnotesize{\sl $^{1}$ University of Padova, \sl $^{2}$ Warsaw University of Technology, \sl $^{3}$ Systems Research Institute, Polish Academy of~Sciences} \\
		\footnotesize{$\ast$ E-mail: antonio.calcagni@unipd.it}
	}

\date{}

\begin{document}

\maketitle

\begin{abstract}
Fuzzy data, prevalent in social sciences and other fields, capture uncertainties arising from subjective evaluations and measurement imprecision. Despite significant advancements in fuzzy statistics, a unified inferential regression-based framework remains undeveloped. Hence, we propose a novel approach for analyzing bounded fuzzy variables within a regression framework. Building on the premise that fuzzy data result from a process analogous to statistical coarsening, we introduce a conditional probabilistic approach that links observed fuzzy statistics (e.g., mode, spread) to the underlying, unobserved statistical model, which depends on external covariates. The inferential problem is addressed using Approximate Bayesian methods, mainly through a Gibbs sampler incorporating a quadratic approximation of the posterior distribution. Simulation studies and applications involving external validations are employed to evaluate the effectiveness of the proposed approach for fuzzy data analysis. By reintegrating fuzzy data analysis into a more traditional statistical framework, this work provides a significant step toward enhancing the interpretability and applicability of fuzzy statistical methods in many applicative contexts.

\end{abstract}

\vspace{2cm}

\section{Introduction}\label{sec1}

The increasing complexity of data in modern contexts is fundamentally transforming the landscape of analysis and modeling, driving the need for more sophisticated approaches and innovative methodologies. This shift, often described as the transition from the curse to the blessing of dimensionality \cite{donoho2000high}, is driven by the increasing prevalence of unstructured and semi-structured data sources. However, this complexity encompasses more than just the variety and volume of data. Instead, it arises from challenges such as unconventional data structures -- like networks and trees -- and the need to address diverse sources of uncertainty, including sampling variability, measurement errors, systematic biases, and instrumental distortions \cite{groves2010total}. Even within more traditional frameworks, complexities persist. A notable example is the use of rating scales to measure subjective attitudes, opinions, or beliefs regarding specific objects. Though longstanding and practical, these scales are influenced by cognitive, affective, and contextual factors inherent to the response process \cite{tourangeau2018survey}. Consider, for instance, the evaluation of sensory attributes, such as the overall experience of tasting a sparkling wine using a 7-point hedonic scale ranging from ``completely dislike'' to ``completely like''. In this scenario, raters rely on long-term memory, activate affective and cognitive components, and integrate them to form and express their final opinions. This decision-making process can introduce uncertainty at various stages, as conflicting cognitive and emotional demands influence the final response. Consequently, the recorded response may not fully capture the underlying experience. To address such complexities, fuzzy rating scales have been developed and employed successfully in several applicative contexts \cite{de2014fuzzy,calcagni2022psychometric,ramos2019applying}. In this setting, fuzzy mathematics provides a valuable framework to represent this type of uncertainty as epistemic imprecision or fuzziness alongside the traditional randomness that drives the sampling process \cite{couso2014statistical}. This approach allows for the simultaneous accommodation of randomness and fuzziness, enhancing the precision and reliability of statistical models. 

However, the use of rating scales is just one example where fuzzy data analysis can be conveniently applied. While it is widely assumed that data from random experiments can be numerically represented, several research contexts yield imprecise data that cannot be adequately expressed on a standard numerical scale. Although some researchers simplify their models to avoid this issue, fuzzy random variables (or random fuzzy numbers) offer a natural way to model both randomness and fuzziness in real-world evaluations. Nevertheless, as in any area of mathematical and statistical modeling, a solid theoretical foundation is crucial. In particular, inference based on random fuzzy numbers needs to be framed within a probabilistic structure, ensuring that methodologies for imprecise data are statistically well-justified.

That said, despite the availability of numerous statistical tools designed to handle fuzzy data, a consistent and comprehensive framework for their statistical analysis is still lacking.  In particular, a general methodology comparable to the Generalized Linear Models (GLMs) framework in traditional statistics has yet to be developed. Although the fuzzy methods developed so far are firmly grounded in measure-theoretic principles \cite{kratschmer2001unified, gil2006overview}, their practical use is limited by the absence of general operational parametric probability models for fuzzy random variables. Furthermore, the lack of a unified inferential framework that provides broadly applicable asymptotic results adds to the challenges of their implementation \cite{gonzalez2009simulation}.

Based on the general premise that fuzzy data arise from a process analogous to statistical coarsening \cite{gill1997coarsening,cattaneo2017likelihood,nguyen2006random}, this contribution introduces a comprehensive method that incorporates the epistemic mechanism assumed to underlie the generation of fuzzy data. The proposed approach utilizes a conditional probabilistic framework to connect the observed characteristics of fuzzy numbers (e.g., mode or spread) to the true, yet unobserved, statistical model underlying the random sample prior to fuzzification. Estimation and inference are performed using the Gibbs sampler-based approach, wherein the full conditional distribution is approximated through sampling from a quadratic approximation of the target posterior distribution. Notably, our solution adopts a regression-based framework similar to GLMs, allowing for incorporating covariates into the modeling process. As this work represents the first attempt to provide such a generalization, we restrict our attention to the case of bounded fuzzy variables only, which are frequently encountered in social science studies \cite{smithson2006fuzzy}. 

The main findings presented in this paper can be summarized as follows:
\begin{enumerate}
	\item[(i)] An epistemic-oriented two-stage generation scheme for fuzzy samples is introduced.
	This approach, closely related to a GLM-like hierarchical Bayesian framework, also enables a model-oriented defuzzification procedure. 
	As a result, the underlying parameters of the statistical model can be estimated even when the data are fuzzy or imprecise and obscure their actual values.
	Although we focus on the case of bounded fuzzy sets represented by Beta-type fuzzy numbers, the proposed approach can also be generalized to more complex cases involving unbounded cases.
	\item[(ii)] Analytical formulas are provided for the Bayesian formulation of the proposed sampling mechanism.
	To numerically solve these, we employ a quadratic approximation and a multivariate Skew-Normal distribution-based approximation, both used in conjunction with a Gibbs sampling procedure.
	Our numerical results show that the quadratic approximation yields low errors and provides high-quality estimates.
	\item[(iii)] To demonstrate the practical usefulness of the overall approach (i.e., the sampling scheme and the accompanying numerical procedures), four case studies based on real-life fuzzy data are analyzed. The performance of the proposed method is assessed using posterior predictive checks based on numerical simulations for each dataset. The final case study additionally involves parameter estimation of the underlying model and a detailed analysis of the associated variables. Finally, the Supplementary Materials provide extended numerical and graphical results for the fuzzy data analysis, including a comprehensive comparison with a selection of state-of-the-art fuzzy regression methods and a detailed and illustrated procedure for assessing model misspecification through posterior predictive analyses.
\end{enumerate}

The paper is organized as follows. Section \ref{sec2} provides foundational definitions and key concepts related to fuzzy numbers. Section \ref{sec3} describes the problem under consideration and introduces the proposed solution. Section \ref{sec4} elaborates on the derivation of the Gibbs sampler, while Section \ref{sec5} presents the results of a simulation study assessing the approximations utilized within the Gibbs sampling framework. Section \ref{sec6} evaluates the external validity of the model by testing its ability to replicate externally collected fuzzy data. Finally, Section \ref{sec7} concludes with a summary of the main findings and their broader implications. The data and algorithms supporting this study's findings are openly available at \texttt{https://github.com/antcalcagni/fuzzyGibbs}.

\section{Preliminaries}\label{sec2}

This section provides an overview of the basic concepts and properties of different types of fuzzy numbers. We refer the reader to \cite{DuboisPrade} for more comprehensive details.

%\begin{definition} 
%\label{D} 
A \textit{fuzzy number} (abbreviated further by FN) is an imprecise value characterized by a mapping $\widetilde{A}:\mathbb{R}\to [0,1]$, called a membership function, such that its $\alpha$-cut defined by
\begin{equation}
\widetilde{A}_{\alpha}=
\begin{cases}
\{x\in\mathbb{R}:\widetilde{A}(x)\geq\alpha\} & \text{if}\quad \alpha\in (0,1], \\
cl\{x\in\mathbb{R}:\widetilde{A}(x)>0\} & \text{if}\quad \alpha=0,
\end{cases} \label{eq_acut}
\end{equation}
is a nonempty compact interval for each $\alpha\in [0,1]$, where $cl$ denotes the closure.
%\end{definition} 

Thus, a FN can be described by its membership function $\widetilde{A}(x)$ or equivalently by a family of its $\alpha$-cuts $\{\widetilde{A}_{\alpha}\}_{\alpha\in [0,1]}$.
Two $\alpha$-cuts are of particular interest: the \textit{core} ($\widetilde{A}_1 = \mathrm{core}(\widetilde{A})$), which contains all values fully compatible with the concept modeled by $\widetilde{A}$, and the \textit{support} ($\widetilde{A}_0 = \mathrm{supp}(\widetilde{A})$), which includes values compatible to some extent with the concept $\widetilde{A}$.
The family of all FNs will be denoted by $\mathbb{F}(\mathbb{R})$.

Various types of fuzzy numbers have been explored extensively in the literature. Among these, LR fuzzy numbers (LRFNs) stand out as one of the most commonly utilized classes. We say that a fuzzy number $\widetilde{A}$ is classified as an LRFN if its membership function can be expressed using the following formula
\begin{equation} 
\widetilde{A}(x)=
\begin{cases}
 0 & \text{if}\quad x < a_1,  \\
 L \left( \frac{x-a_1}{a_2 - a_1}\right) & \text{if}\quad a_1 \leq x < a_2 ,  \\
 1 & \text{if}\quad a_2 \leq x < a_3 , \\
 R \left( \frac{a_4 - x}{a_4 - a_3}\right) & \text{if}\quad a_3 \leq x < a_4 , \\
 0 & \text{if}\quad x \geq a_4,  
\end{cases}
\label{eq:LFfn}
\end{equation} 
where $L, R: [0,1] \rightarrow [0,1]$ are continuous and strictly increasing functions satisfying $L(0)=R(0)=0, L(1)=R(1)=1$, and $a_1, a_2, a_3, a_4 \in \mathbb{R}$ such that $a_1\leq a_2\leq a_3\leq a_4$.
Specifically, if  L  and  R  are linear functions defined as
\begin{equation}
	L(x)=\frac{x-a_1}{a_2 - a_1} , \quad R(x)=\frac{a_4 - x}{a_4 - a_3},
\end{equation}
the resulting fuzzy number is called a trapezoidal fuzzy number (TPFN). Furthermore, if  $a_2 = a_3$, the corresponding fuzzy number is called a \textit{triangular fuzzy number} (TRFN).

A specific case of generalized trapezoidal fuzzy numbers (TRFNs) is referred to as beta fuzzy numbers (BFNs) \cite{BAKLOUTI2018259,calcagni2022modeling}. The membership function of a BFN is expressed as
\begin{equation}
    \widetilde{A}(x)=
    \begin{cases}
     0 & \text{if}\quad x < lb,  \\
     \left( \frac{x-lb}{m - lb}\right)^p \left( \frac{ub-x}{ub-m}\right)^q & \text{if}\quad lb \leq x < ub , \\
     0 & \text{if}\quad x \geq ub,  
    \end{cases}
    \label{eq:BFN}
\end{equation}
where $p, q, lb, ub  \in \mathbb{R} $ such that $p, q >0, lb < ub$, and $m$ is the mode of the BFN, defined as
\begin{equation}
    m = \frac{p \; ub + q \; lb}{p+q} .
\end{equation}
Additionally, a BFN can be represented using its mode  $m$  and precision  $s$. In this case, assuming for simplicity $lb = 0$  and  $ub = 1$, the membership function becomes
\begin{equation}
    \tilde{A}(x)= \frac{1}{C} x^{p-1} (1-x)^{q-1} ,
\end{equation}
where
\begin{equation}
    p = 1 +s m, q = 1 + s(1-m) , 
\end{equation}
and
\begin{equation}
  C = \left( \frac{p-1}{p+q-2}\right)^{p-1} \left( 1- \frac{p-1}{p+q-2}\right)^{q-1} .
\end{equation}
This formulation provides a concise and flexible way to model BFNs, simplifying their application in further analyses. 

\begin{figure}[htbp]
\centering
	\hspace*{4.25cm}
	%\resizebox{13cm}{!}{\input{fig1.tex}}
	\includegraphics[scale=0.8]{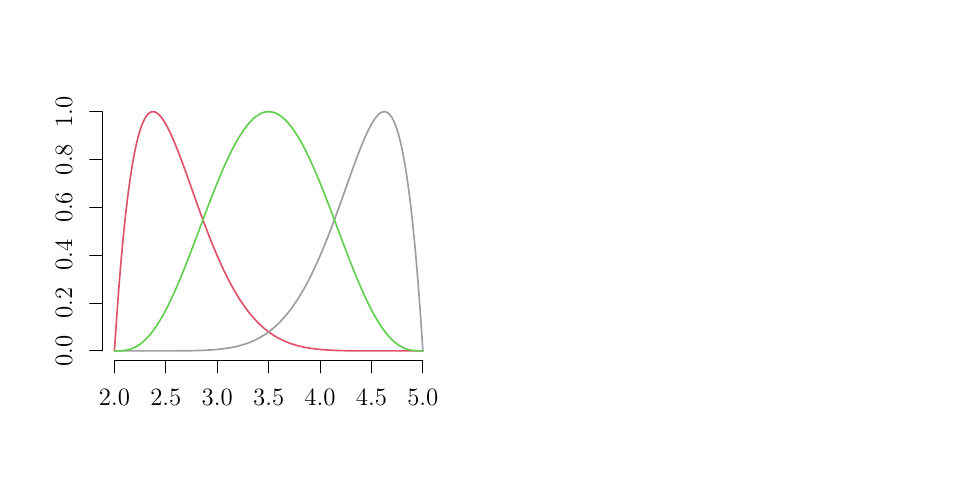}
	\caption{Some examples of (bounded) Beta fuzzy sets.}
	\label{fig1}
\end{figure}

Figure \ref{fig1} illustrates an example of a BFN. Notably, similar to the general case of beta and triangular probability density functions \cite{gupta2004handbook}, the beta function used in the membership function of a BFN can approximate other functions, such as the triangular or Gaussian functions. This is particularly true when the mode  $m$  of the BFN aligns with the mode of the corresponding triangular or Gaussian fuzzy sets.

One of the primary purposes of using fuzzy numbers (FNs) is to model imprecise data, i.e., data samples that cannot be described precisely. To incorporate the random component often encountered in real-life applications, two main approaches to \textit{random fuzzy sets} (or f\textit{uzzy random variables}) have been introduced in the literature. The first approach, commonly called the \textit{ontic view} \cite{Puri}, focuses on modeling inherently random processes with imprecise states. The second approach, known as the \textit{epistemic view} \cite{Kruse1982, Kwakernaak}, stems from incomplete knowledge about the random variable and is associated with the following definition. Given a probability space $(\Omega,\mathcal{F},P)$, a mapping $\widetilde{X}:\Omega \to\mathbb{F}(\mathbb{R})$ is called a \textit{fuzzy random variable} if for each $\alpha\in [0,1]$ $(\inf \widetilde{X}_\alpha):\Omega\to\mathbb{R}$ and $(\sup \widetilde{X}_\alpha):\Omega\to\mathbb{R}$ are real-valued random variables on $(\Omega,\mathcal{F},P)$.

%\begin{definition}
	%Given a probability space $(\Omega,\mathcal{F},P)$, a mapping $\widetilde{X}:\Omega \to\mathbb{F}(\mathbb{R})$ is called a \textit{fuzzy random variable} if for each $\alpha\in [0,1]$ $(\inf \widetilde{X}_\alpha):\Omega\to\mathbb{R}$ and $(\sup \widetilde{X}_\alpha):\Omega\to\mathbb{R}$ are real-valued random variables on $(\Omega,\mathcal{F},P)$.
%\end{definition}

Thus, a fuzzy random variable $\widetilde{X}$ can be understood as an \textit{imperfect perception} of a ``classical'' (i.e., real-valued but unobserved) random variable $X$, often referred to as the \textit{original} of $\widetilde{X}$. Consequently, a fuzzy random sample $\widetilde{\mathbb{X}} = (\widetilde{X}_1, \ldots, \widetilde{X}_n)$ represents a fuzzy perception of a standard real-valued random sample. This interpretation aligns with the well-established modeling of coarsening mechanisms in traditional statistical frameworks \cite{gill1997coarsening}.

%%%%%%%%%%%%%%%%%%%%%%%%%%%%%%%%%%%%%%%%%%%%%%%%%%%%%%%%%%%%%%%%%%%%%%%%%%%%%

\section{A novel conditional sampling schema}\label{sec3}

\subsection{Statement of the problem}\label{sec3.1}

Throughout this paper, we denote by $\mathbf{X}$ an $n \times J$ matrix of covariates (or predictors). The response variable to be modeled is represented by the $n \times 1$ observed random vector $\mathbf{y}$, where $y_i$ refers to its generic entry. The fuzzy counterparts are denoted by $\mathbf{\widetilde{y}}$ and $\widetilde{y}_i$, respectively. To simplify notation, we use lowercase and uppercase letters interchangeably to represent random variables and their realizations, with the distinction evident from the context. Following the standard GLM-like framework, the model includes a systematic component $\eta = g(\Exp{Y}) = \mathbf X\boldsymbol\beta$, where $\boldsymbol{\beta}$ represents the vector of regression coefficients. Additionally, the scalar parameter $\phi$ describes the model's dispersion (or precision), an element of the variance function $\Var{Y} = h(\Exp{Y}\phi)$, for which no regression structure is assumed. 

We start with the general statement of the problem.
Let $\mathbb Y = \{Y_1,\ldots,Y_n\}$ denote a set of $n$ independent continuous random variables and let $\mathbf{\widetilde y} = (\widetilde y_1,\ldots,\widetilde y_n)$ be a sample of fuzzy observations. Due to epistemic or post-sampling uncertainty-based processes \cite{cao2024novel}, the vector $\mathbf{\widetilde{y}}$ serves as a blurred representation of the true, yet unobserved, vector of crisp random outcomes $\mathbf{y}$. The goal is to study the joint density $f_{Y_1, \ldots, Y_n}(\mathbf{y}; \boldsymbol{\theta}_y)$ in order to make inference about the parameter vector $\boldsymbol{\theta}_y$, given the fuzzy sample $\mathbf{\widetilde{y}}$. From the epistemic perspective, fuzzy observations are regarded as stochastic outcomes influenced by a combination of non-random and systematic uncertainty, which obscure the true but unknown realizations $\mathbf{y}$.

In this contribution, we apply a specific type of blur to the true values $\mathbf{y}$, motivated by the epistemic interpretation of fuzzy random variables discussed above. We assume that the fuzzy observations $\mathbf{\widetilde y}$ are described by Beta-type fuzzy numbers (see Sect.~\ref{sec3.2}), so that each Beta-type fuzzy observation can be conveniently characterized by a mode and a precision parameter, expressed as $\widetilde{y}_i = \{m_i, s_i\}$.

\subsection{Proposed solution}\label{sec3.2}

The proposed approach is based on formalizing the two-stage process underlying the generation of fuzzy numbers. Given the inherent interaction between these stages, our solution adopts a conditional probabilistic framework. This framework establishes a connection between the parameters of fuzzy numbers -- such as modes, spreads, and, where applicable, membership function characteristics -- and the random outcomes of the model $f_Y(y;\boldsymbol{\theta})$.

\begin{figure}[!t]
	\centering
	%\resizebox{9cm}{!}{\input{fig2.tex}}
	\includegraphics[scale=0.5]{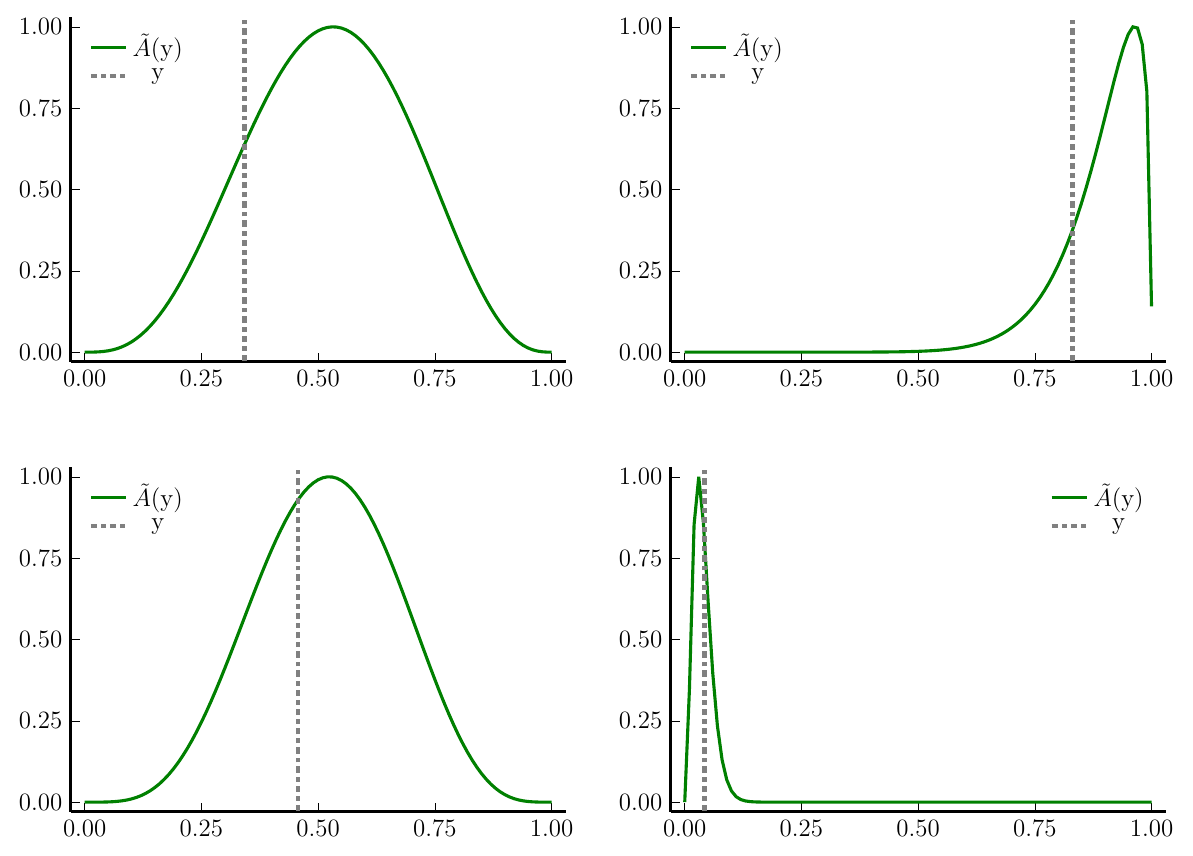}
	\caption{Examples of Beta-type fuzzy numbers $\widetilde{A}({y})$ (green curves) blurring the true unobserved realizations $y$ (dashed vertical gray lines).}
	\label{fig2}
\end{figure}

For Beta-type fuzzy numbers, the two-stage sampling process involving fuzzy numbers with modes $m_i$ and precision $s_i$ can be structured as follows:
\begin{align}
	&y_i \sim f_{Y}(y;\boldsymbol{\theta}_y), \label{eq1} \\
	&s_i \sim \mathcal{G}(s;\alpha_s,\beta_s), \label{eq2} \\
	&m_i|s_i,y_i \sim  \mathcal{B}_{4P}(m;s_iy_i,s_i-s_iy_i,lb,ub). \label{eq3}
\end{align}
Notably, in \eqref{eq1}, $f_{Y_i}(Y;\boldsymbol{\theta}_y)$ represents the random variable governing the non-fuzzy sampling process, with $\eta_i = g(\Exp{Y_i}) = \mathbf x_i\boldsymbol\beta$. In \eqref{eq2}, the precision (or spread) of the fuzzy number is modeled using a Gamma random variable $\mathcal{G}$, with the shape parameter and scale parameters $\alpha_s > 0$ and $\beta_s > 0$., respectively. Finally,  \eqref{eq3} introduces a 4-parameter Beta conditional random variable, which governs the mode of the fuzzy number as a function of the true yet unobserved outcome $y_i$ and the spread $s_i$. 
Figure \ref{fig2} illustrates the interplay of these components, showing how fuzziness affects the true unobserved random realization by \eqref{eq1}-\eqref{eq3}.

A key feature of the integration process described in \eqref{eq3} is its capacity to avoid the trivial case where the mode $m_i$ of the observed fuzzy number equals the latent realization $y_i$. While this scenario can occur (e.g., see Fig. \ref{fig2}, second panel, counterclockwise), the proposed modeling approach accommodates the broader case where the true unobserved realization lies within a plausible interval. This interval represents partial information about uncertain data captured through a possibility distribution. Such a scenario reflects a typical case where the observer possesses only approximate knowledge of the data and can specify merely a graded interval of plausible values \cite{denoeux2011maximum}. Note that, relying on defuzzified data (e.g., the mode $m_i$ or center-of-gravity index) instead of the observed fuzzy sets as input for inference on $\boldsymbol{\theta}_y$ would lead to biased estimates, particularly for variance parameters, as shown in prior studies \cite{calcagni2022modeling}.

To better understand the rationale behind the proposed conditional sampling schema, particularly the interaction of its three terms in shaping the expectation and variance of the modes of the observed fuzzy numbers, consider as follows (assuming $lb=0$ and $ub=1$ for simplicity):
\begin{align}
	& \Exp{M} = \Exp{Y},\label{eq3a}\\
	& \Var{M} = (\Var{Y}-\Var{Y}c) + \Exp{Y}(1-\Exp{Y})c, \label{eq3b}
\end{align}
where  $c = \mathbb{E}[(S+1)^{-1}] \in (0,1)$ is the expected value of the inverse shifted Gamma distribution \cite{shishter2022modelling}. These expressions show that while the mode of the observed fuzzy set matches the unobserved true realization in expectation, its variance reflects a combined influence of the variability of the true random variable  $Y$  and the degree of fuzziness  $S$.

Consider $S \sim \mathcal{G}(s; 1, \beta_s)$ in the shape-scale parameterization for simplicity. It is evident that as  $\beta_s \to 0$ -- a scenario representing high fuzziness in the observed fuzzy numbers -- the scaling factor  $c \to 1$, leading to  $\Var{M} < \Exp{Y} $. Conversely, as $ \beta_s \to \infty$ -- representing progressively reduced fuzziness -- the scaling factor  $c \to 0$, and  $\Var{M} \approx \Var{Y}$ .

When  $c \geq 0.5$, which occurs at  $\beta_s = 1.639$, the dispersion of the observed modes becomes significantly smaller than that of the unobserved crisp realizations. This happens because  $c$  attenuates the variance of  $Y$, leading to a phenomenon akin to underdispersion \cite{sellers2017underdispersion}. This underdispersion can introduce bias when estimating  $\Var{Y}$  based on fuzzy observations. To mitigate this issue, cases where  $c \ll 0.5$  are preferable -- for instance, at  $\beta_s = 10.121$,  $c = 0.25$, ensure the observed modes' variance closely approximates its counterpart for the unobserved realizations.

Overall, the random variable  $S$, which governs the degree of fuzziness, introduces variability that blurs the relationship between  $M$ and $Y$  by attenuating  $\Var{Y}$. This effect becomes more pronounced with high variability in  $S$, diminishing the ability of  $M$  to fully represent the variability of  $Y$. For completeness, derivations of \eqref{eq3a}-\eqref{eq3b} are provided in the Appendix.

\subsection{Remarks and comments}\label{sec3.3}

\subsubsection{Sampling process}
Similar to statistical coarsening, the sampling process is modeled as a two-stage mechanism: first, a random experiment is conducted, followed by the fuzzification of the observed outcomes. In this framework,  $\widetilde{A}(Y_i = y) \in [0,1]$  represents a soft constraint indicating the possibility that  $Y_i$  takes the value  $y$. An illustrative example of this dual-stage mechanism is a classification task involving chocolate types, distinguished based on aromatic compounds (e.g., pyrazine, aldehyde) and categorized by aroma intensity using a fuzzy variable with four categories (e.g., ``intense'', ``enveloping'', ``bold'', ``prodigious''). Counting how many chocolate bars fall into each category yields fuzzy counts -- a collection of fuzzy natural numbers. These counts encapsulate the inherent variability in chocolate types, driven by the stochastic nature of the \textit{sampling process},  $Y_i \sim f_Y(y; \boldsymbol{\theta}) $, and the subjective nature of aroma intensity evaluations, represented by fuzzy sets  $\widetilde{A}(Y_i = y)$  over the support of  $Y_i$, derived from the classification procedure.

\subsubsection{Assumption $y_i \indep s_i$}
While this assumption might initially seem overly simplistic, it is entirely consistent with the two-stage interpretation of the sampling process. In this framework, epistemic fuzziness reflects the observer’s state of knowledge and is linked to measurement mechanisms that occur after the sampling process is complete. From this perspective, epistemic fuzziness can be viewed as a form of systematic error stemming from limitations in the measurement process, which could potentially be mitigated by acquiring additional information about  $Y_i$. Although epistemic fuzziness may depend on relevant covariates (e.g., level of expertise: a group of experts may assess experimental outcomes with greater confidence than non-experts), it remains a secondary factor within the broader sampling process. For further insights into post-sampling measurement effects, see \cite{cao2024novel}.

\subsubsection{The choice of Gamma and Beta random variables}
Although various options could have been chosen to represent the precision \eqref{eq2} and mode \eqref{eq3} components, the Gamma and Beta distributions offer several advantageous properties. The Gamma distribution, for example, ensures positivity for the precision component while maintaining computational efficiency. Similarly, the four-parameter Beta distribution is selected for its flexibility and ability to capture potential skewness introduced by the fuzziness mechanism. Its log-concavity and unimodality further simplify the computations required during the estimation process (see Section \ref{sec4}).

\subsubsection{The model $f_Y(\mathbf y;\boldsymbol\theta_y)$}
The conditional model is typically designed for fuzzy bounded data. In this case, several statistical models can be applied, including the Beta \cite{ferrari2004beta}, Kumaraswamy \cite{mitnik2013kumaraswamy}, Logitnormal \cite{pinson2012very}, and Logbilai \cite{pena2023unit} distributions. However, when the true underlying model is unbounded, such as for the Gaussian case, the fuzzification process still bounds the actual unobserved realizations  $\{y_1, \ldots, y_n\}$ within the set of most plausible values.

For unbounded models  $f(\mathbf y;\boldsymbol\theta_y)$, the bounding statistics $lb$ and $ub$  can reasonably be approximated by computing the infimum and supremum of the sets  $\mathbb{A}_0^{-} = \{\inf(\widetilde{A}_0^{(1)}), \ldots, \inf(\widetilde{A}_0^{(n)})\}$  and  $\mathbb{A}_0^{+} = \{\sup(\widetilde{A}_0^{(1)}), \ldots, \sup(\widetilde{A}_0^{(n)})\}$, respectively, derived from the supports of the observed fuzzy numbers.

\subsubsection{Assessing model misspecifications}
One can rely on posterior predictive checks and the examination of marginal posterior distributions to assess whether the assumed two-stage fuzzification mechanism adequately captures the data-generating process. In particular, model misspecifications -- such as those arising from more complex blurring mechanisms involving additional stages -- can be detected by evaluating discrepancies between observed and model-implied predictions. Key indicators include poor coverage of the supports and fuzziness, increased posterior variance, and systematic bias in predictive summary statistics. These diagnostics offer insight into whether the model assumptions are violated, especially in the presence of structurally biased fuzzy data, such as $\delta$-shifted observations that allow for fuzzy data not containing the true unobserved realizations. Further details and examples are provided in Section 3 of the Supplementary Materials.

\subsubsection{Model implications in real contexts}
It should be highlighted that our proposal is suitable for data collected in contexts where fuzziness reflects partial knowledge, vague measurement, or perceptual ambiguity -- such as expert assessments, rating tasks, and decision support systems. However, this proposal does not aim to model the internal cognitive mechanisms behind fuzzification. Instead, more simply, it offers a statistical approximation of epistemic processes, allowing practitioners to analyze imprecise data without assuming that respondents consciously implement the underlying probabilistic model. Because of this, existing fuzzy rating procedures, like those proposed by \cite{de2014fuzzy,calcagni2022psychometric}, naturally align with the assumptions of the conditional model and may be seen as compatible backend mechanisms underlying the observed fuzzy data.

That said, in this context, the core of a fuzzy number $\tilde{A}_1^{(i)}$ is identified with the mode $m_i$, representing the value most compatible with the underlying latent variable $Y_i$, such as a rater's true but unobserved judgment. The support $\tilde{A}_0^{(i)}$ is governed by the spread parameter $s_i$, which controls the decay of the membership function around the mode. Therefore, the core captures the most plausible value, while the support reflects a broader range of less precise, yet still compatible, values.

\section{Inference on $\boldsymbol\theta_y$}\label{sec4}

\subsection{A general idea}

Given the conditional sampling structure governing the fuzzification process of  $y_1, \ldots, y_n$, a natural approach for inferring $\boldsymbol{\theta}_y$ involves a deblurring procedure. This method leverages the observed values $\mathbf{\widetilde{y}}$ as surrogates for the unobserved realizations $\mathbf{y}$, thereby facilitating the use of the Gibbs sampling algorithm.

Formally, let the observed modes and precisions be denoted as $\mathbf{D} = (\{m_1, \ldots, m_n\}, \{s_1, \ldots, s_n\})$. This framework allows for sampling from the posterior distribution $\pi(\boldsymbol{\theta}_y, \mathbf{y} \mid \mathbf{D})$ by iteratively drawing samples from the conditional posterior distributions, as follows (for $t > 1$):
\begin{align}
	& \mathbf y^{(t)} \sim \pi(\mathbf y|\mathbf D, \boldsymbol{\theta}_y^{(t-1)}), \label{eq4a} \\
	& \boldsymbol{\theta}_y^{(t)} \sim \pi(\boldsymbol{\theta}_y|\mathbf D, \mathbf{{y}}^{(t)}). \label{eq4b}
\end{align}
The desired distributions are derived by appropriately re-arranging  the joint distribution as follows
\begin{align*}
	f(\mathbf D,\mathbf y,\boldsymbol{\theta}) & = f(\mathbf D|\mathbf y,\boldsymbol{\theta})f(\mathbf y|\boldsymbol{\theta})f(\boldsymbol\theta)\nonumber\\
	& = f(\mathbf m|\mathbf y,\boldsymbol{\theta}_y)f(\mathbf s|\boldsymbol{\theta}_s)f(\mathbf y|\boldsymbol{\theta}_y)f(\boldsymbol\theta_y)f(\boldsymbol\theta_s),
\end{align*}
where $f(\boldsymbol{\theta})$ represents the prior distribution over the model parameters.

It is worth noting that the Gibbs sampler simplifies significantly due to the independence between  $Y_i$ and $S_i$. In particular, the parameters $\boldsymbol{\theta}_s = \{\alpha_s, \beta_s\}$, as defined in \eqref{eq3}, can be straightforwardly estimated via maximum likelihood from $\mathbf{s}$. This simplification arises because marginalizing over $\boldsymbol{\theta}_s$ collapses the term $f(\mathbf{s} \mid \boldsymbol{\theta}_s)f(\boldsymbol{\theta}_s)$, effectively decoupling it from the remaining terms in the joint distribution. It is also worth highlighting that \eqref{eq1}-\eqref{eq3} are quite general and, in this context, do not naturally lend themselves to analysis within a conjugate Bayesian framework. As a result, sampling from the conditional posteriors $\pi(\mathbf{y} \mid \boldsymbol{\theta}_y, \mathbf{D})$ and $\pi(\boldsymbol{\theta}_y \mid \mathbf{y}, \mathbf{D})$ can present challenges. To address these difficulties, we adopted a fully approximated solution in which both steps are performed using posterior approximation techniques. The details of this approach are thoroughly described in Sections \ref{sec4.1}-\ref{sec4.2}.

\subsection{Approximation of $\pi(y|\boldsymbol\theta_y,\mathbf D)$}\label{sec4.1}

The quadratic approximation is derived by aligning the parameters of the log posterior density $\log \pi(y|\ldots)$ with the first- and second-order derivatives of a chosen proposal distribution. The selection of this proposal distribution is guided by its suitability to represent the unnormalized posterior target \cite{tierney1986accurate,tanner1993tools}. In our case, the proposal posterior density is modeled as a univariate four-parameter Beta distribution, characterized by its mode $\lambda \in (lb,ub)$ and precision $\sigma \in \mathbb{R}^+$. More formally, let in general $z^* = (z-lb)/(ub-lb)$ and $z^\dagger = lb+(ub-lb)z$ and fix the observation $i\in\{1,\ldots,n\}$. Then, we obtain
\begin{align}
\ln \pi(y|\boldsymbol{\theta}_y,\ldots) & \propto - {\ln\Gamma(y^*s) -\ln\Gamma(s-sy^*) + sy^* \ln m^*} + \ln f_{Y}(y;\boldsymbol{\theta}_y) \nonumber \\
	& \propto h(y;m^*,s) + \ln f_{Y}(y|\boldsymbol{\theta}_y) \nonumber \\
	& {\approxx \ln \mathcal{B}_{4P}(y;{\lambda}\sigma,\sigma-{\sigma}\lambda,lb,ub)}.
\label{problem3case1}
\end{align}
The unknown parameters of the proposal distribution are determined by iteratively solving the following equations ($k=1,2$)
\begin{equation}
	\frac{\partial^k}{\partial y^k} \ln \mathcal{B}_{4P}(y;\lambda\sigma,\sigma-\sigma\lambda,lb,ub) = \frac{\partial^k}{\partial y^k} \Big(h(y;m^*,s) + \ln f_Y(y;\boldsymbol{\theta}_y) \Big),
\label{problem3case1DA}
\end{equation}
with a generic recursion involving the steps:
\begin{align}
	& y = \widehat\lambda', \nonumber\\
	& k_1 = \frac{\partial}{\partial y} \ln f_Y(y^\dagger|\boldsymbol\theta_y) + s \ln\left(\frac{m^*}{1 - m^*}\right) + s\left(-\psi^{(0)}(sy) + \psi^{(0)}(s - sy)\right), \nonumber\\[0.1cm]
	& k_2 = \frac{\partial^2}{\partial y^2} \ln f_Y(y^\dagger|\boldsymbol\theta_y) - s^2\left(\psi^{(1)}(sy) + \psi^{(1)}(s - sy)\right), \nonumber \\[0.1cm]
	& \widehat\lambda = \frac{1 + y(-2 + k_1 + \widehat\sigma' - k_1 y)}{\widehat\sigma'},\\[0.1cm]
	& \widehat\sigma = \frac{1 - (-1 + y)y(-2 + k_2(-1 + y)y)}{\widehat\lambda' - 2\widehat\lambda' y + y^2}.
\end{align}

The recursion continues until  $\left| \frac{y}{\widehat{\lambda}} - 1 \right| < \epsilon$, where $\epsilon$ is a small positive threshold. Here, $\psi^{(m)}$ stands for the polygamma function of order $m$, and $\{\widehat{\lambda}{\prime}, \widehat{\sigma}{\prime}\}$ denote the parameter values from the previous iteration. Finally, the solutions $\widehat\lambda$ and $\widehat\sigma$ are utilized in the initial Gibbs step to draw samples from the (approximated) conditional posterior distribution $y^{(t)}_i \sim \mathcal{B}_{4P}(y;{\widehat\lambda}\widehat\sigma,\widehat\sigma-{\widehat\sigma}\widehat\lambda, lb,ub)$.

It should be emphasized that the selection of the four-parameter Beta distribution, among many other possible distributions, is driven by its exceptional flexibility in accommodating both symmetric and asymmetric cases \cite{gupta2004handbook}. In our specific context, this choice is particularly justified because, when $y$ approaches its boundary values, the fuzzification process accounts for this constraint. This advantage would not be realized if the same probabilistic model for $Y$ were applied instead. Although such an approach would result in a more straightforward form of the posterior distribution $\pi(y | \boldsymbol\theta_y, \mathbf{D})$, simplifying the computational requirements, it would lack the flexibility inherent to the Beta model. Furthermore, the use of the Beta distribution is reinforced by the observation that the function $h(y; m^, s)$ closely resembles the kernel of the log-Beta distribution evaluated at $m^*$, preserving the property of log-concavity required for an accurate second-order approximation.

\subsection{Approximation of $\pi(\boldsymbol\theta_y|\mathbf y,\mathbf D)$}\label{sec4.2}

The approximation of the remaining term of the Gibbs sampler is based on the multivariate Skew-Normal distribution \cite{zhou2024tractable}: 
\begin{equation}
	\ln \pi(\boldsymbol\theta_y|\mathbf y,\mathbf D) \approxx \ln \mathcal{SN}(\boldsymbol\theta_y;\boldsymbol\mu,\boldsymbol\Sigma,\boldsymbol\delta).
\end{equation}

While the simpler multivariate Normal distribution could be used in this context -- since it matches, up to a second-order approximation, the mean $\boldsymbol\mu$ and variance $\boldsymbol\Sigma$ with those of the log-posterior distribution—the adoption of an asymmetric distribution allows for handling skewed posterior distributions, albeit at the cost of requiring a third-order approximation through the parameter $\boldsymbol\delta$. In our problem, we match the mode, the negative Hessian at the mode, and the third-order unmixed derivatives at the mode of the observed log-posterior $\ln \pi(\boldsymbol\theta_y|\mathbf y,\mathbf D)$ to the corresponding parameters of the multivariate Skew-Normal distribution, thereby forming a third-order approximation of the original distribution. As in the previous scenario, the solutions are computed numerically using iterative solvers. We direct the reader to Algorithm 2 in \cite{zhou2024tractable} for a detailed explanation of the numerical solutions. Additional insights into Skew-Normal approximations in Bayesian settings can be found in \cite{zhou2024tractable, durante2024skewed, pozza2024skew}.

\section{Simulation study}\label{sec5}

In this section, we present the findings of a brief simulation study evaluating the quadratic posterior approximation of $\pi(y|\boldsymbol{\theta}_y,\mathbf D)$ using the $\mathcal{B}_{4P}$ approximation. Given that approximated Bayesian methods are well-established in the literature, we focus exclusively on the novel aspects of our approach. For details on the convergence properties of the approximation method used in this study, we refer the reader to \cite{miller2019fast,zhou2022bayesian,kawakami2023approximate}.

\begin{table}[hbt]
\centering
\caption{Distributions applied in the simulation study}\label{tab100}
\begin{tabular}{c|c|c}
\hline 
 Model & Distribution & Values of the parameters   \\ 
\hline
$\mathcal{G}(s;\alpha_s,\beta_s)$ & $\mathcal{G}(\alpha,\beta)$ & $\alpha \in \{15,30,45\}, \beta \in \{ 5,15,35\}$\\
\hline
\multirow{6}{*}{$f_Y(y;\boldsymbol{\theta}_y)$} & $\mathcal{LGN} (\mu, \sigma)$ &  $ \mu \in \{-1.85,0,1.85\} , \sigma \in \{ 1,2,3.5\} $\\
& $\mathcal{B}(a,b)$ & $a \in \{0.5,2,5\}, b \in \{ 0.5,1,3\}$ \\
& $\mathcal{LBL}(\theta)$ & $\theta \in \{0.3,0.5,1,1.5\}$ \\
& $\mathcal{K}(a,b)$ &  $a \in \{0.5,2,5\}, b \in \{ 0.5,3,6\}$  \\
& \multirow{2}{*}{$\mathcal{TN} (a,b,\mu, \sigma)$}  & $ a=0, b=1, \mu \in \{ 0.2,0.3,0.5\}, $  \\
& & $\sigma \in \{ 0.1,0.17,0.2 \}$ \\
\hline
\end{tabular}
\end{table}

To evaluate the quality of the proposed derivative-based approximation (DA) using the four-parameter Beta distribution $\mathcal{B}_{4P}$, we considered a set of various distributions for $f_Y(y;\boldsymbol{\theta}_y)$ with differing parameter configurations for this density and $\mathcal{G}(s;\alpha_s,\beta_s)$, as applied in model \eqref{eq2}. For simplicity, we set $lb=0$ and $ub=1$. Table \ref{tab100} summarizes the considered distributions: $\mathcal{G}(\alpha,\beta)$ represents the Gamma distribution with shape $\alpha$ and scale $\beta$; $\mathcal{LGN}(\mu, \sigma)$ denotes the logit-normal distribution with location $\mu$ and scale $\sigma$; $\mathcal{B}(a,b)$ corresponds to the Beta distribution with shape parameters $a$ and $b$; $\mathcal{LBL}(\theta)$ refers to the log-Bilal distribution \cite{Altun2021-to} with parameter $\theta$; $\mathcal{K}(a,b)$ stands for the Kumaraswamy distribution \cite{FLETCHER1996259} with parameters $a$ and $b$; and $\mathcal{TN}(a,b,\mu, \sigma)$ describes the normal distribution truncated to the interval $(a,b)$, with mean $\mu$ and standard deviation $\sigma$. We computed the total variation and Hellinger distances between the target distribution $\pi(y|\boldsymbol{\theta}y,\mathbf D)$ and its approximation using $\mathcal{B}{4P}$. To reduce randomness and ensure a comprehensive exploration of the interval $[0,1]$, each numerical experiment was repeated 500 times. The main results are presented below, with extended outputs provided in the Supplementary Materials.

\begin{figure}[htbp]
  \centering
	%\resizebox{12cm}{!}{\input{fig9.tex}}
	\includegraphics[scale=0.7]{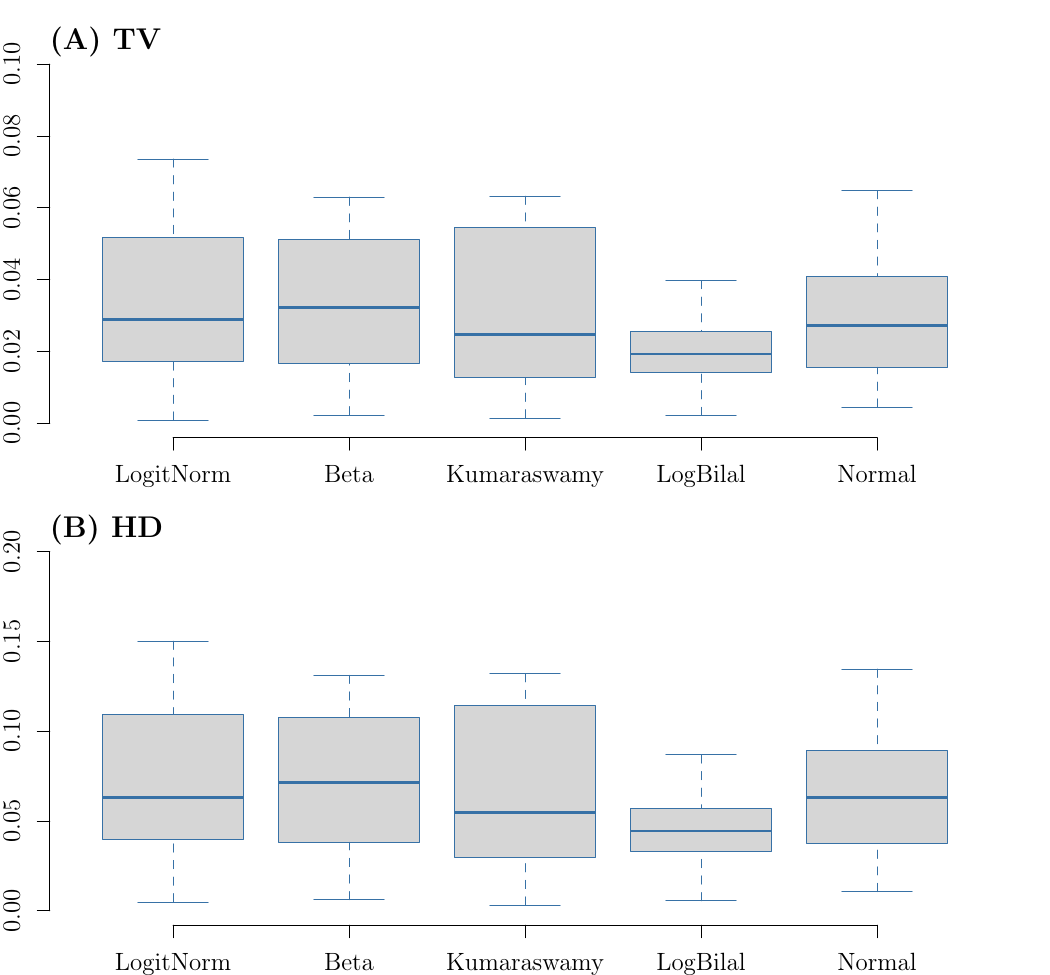}
	\caption{Boxplots of the estimated total variation (TV) and Hellinger (HD) distances in the DA analysis for $\pi(y|\boldsymbol{\theta}_y,\mathbf{D})$}
	\label{fig100}
\end{figure}

As shown in the boxplots in Fig. \ref{fig100} and the summary of means and standard errors in Table \ref{tab101}, the DA approach proposed in \eqref{problem3case1DA} demonstrates high quality. To enhance readability, individual outliers were excluded from the graphs. Typically, the medians of the estimated differences in approximation are around 0.02–0.03 for the total variation distance and 0.04–0.07 for the Hellinger distance. Similarly, the means of the differences range from 0.02 to 0.34 for the total variation distance and from 0.045 to 0.073 for the Hellinger distance, while the standard errors remain very small, approximately 0.0001–0.0002.

\begin{table}[hbtp]
\centering
\caption{Estimated means and standard errors for the total variation (TV) and Hellinger (HD) distances in the simulation study}\label{tab101}
\begin{tabular}{c|cc|cc}
\hline 
 Distribution & \multicolumn{2}{c|}{TV} & \multicolumn{2}{c}{HD}   \\ 
\hline
& Mean & Std. error & Mean & Std. error \\
\hline
$\mathcal{LGN} (\mu, \sigma)$ & 0.0339 & 0.0001 & 0.0731 & 0.0002 \\
$\mathcal{B}(\alpha,\beta)$ & 0.0336 & 0.0001 & 0.0728 & 0.0002 \\
$\mathcal{LBL}(\theta)$ & 0.02 & 0.0001 & 0.0453 & 0.0001  \\
$\mathcal{K}(a,b)$ & 0.0315 & 0.0001 & 0.0679 & 0.0002 \\
$\mathcal{TN} (a,b,\mu, \sigma)$ & 0.0285 & 0.0001 & 0.064 &  0.0002 \\
\hline
\end{tabular}
\end{table}

\section{Applications}\label{sec6}

In this section, we illustrate the application of the proposed model using a collection of pre-existing datasets. The section is structured into four subsections: the first three assess the model’s external validity through a Posterior Predictive Check (PPC) approach, while the final subsection presents a complete case study.

In the context of this research, the PPC methodology follows the steps outlined by \cite{gelman1996posterior}:
\begin{enumerate}
	\item[(i)] Define the non-fuzzy model $f_Y(y; \boldsymbol{\theta}_y)$.
	\item[(ii)] Estimate the model parameters $\boldsymbol{\theta}_y$ using the observed fuzzy dataset $\mathbf{\widetilde{y}}_{n\times 1}$ using the approximated Gibbs sampler (see Section \ref{sec4}).
	\item[(iii)] Generate $B$ new instances of fuzzy data $\mathbf{\widetilde{y}}_1, \ldots, \mathbf{\widetilde{y}}_B$ based on the model's conditional structure (see Eqs. \ref{eq1}-\ref{eq3}).
\end{enumerate}

The model’s performance is evaluated by comparing three observed statistics $S(\mathbf{\widetilde{y}})$ -- centroids, supports (i.e., 0-cuts), and fuzziness (measured using Kaufman’s index \cite{kaufmann1975introduction}) -- against the distribution of their simulated counterparts. These comparisons are conducted using interquartile ranges ($Q_3 - Q_1$), 95\% coverage probabilities (CP), and a transformed Bayesian $p$-value (bP), which is adjusted for interpretability similar to the frequentist $p$-value.

For all datasets, the approximate Gibbs sampler described in Section \ref{sec4} is applied, running five independent chains with 4000 samples each, following a burn-in period of 2000 samples. Weakly informative priors are used, ensuring vague yet reasonable constraints on parameter values. Chain convergence is assessed using standard diagnostics, including the Gelman-Rubin statistic ($\widehat{R}$) and Effective Sample Size (ESS) \cite{gelman2008bayesian}. In cases requiring model comparison, the Widely Applicable Information Criterion (WAIC) is employed \cite{vehtari2017practical}.

%%%%%%%%%%%%

\subsection{Reforestation data (Colubi, 2009)}\label{sec6.1}

The first dataset, published by \cite{colubi2009statistical}, originates from a comprehensive study conducted at the Indurot Institute (University of Oviedo) on the progress of a reforestation project in Asturias (Spain). A key aspect of the study was assessing tree quality, which was evaluated based on subjective expert judgments rather than a single real-valued magnitude. Experts considered multiple factors, including leaf structure, root system, and height-to-diameter ratio.

A fuzzy-valued scale was adopted since a standard categorical scale failed to capture the nuances of these expert perceptions. The scale’s support was expressed as a percentage, with 0 indicating no quality and 1 representing perfect quality. For our analysis, we used a subsample of $n=200$ fuzzy observations. The three tree species (Betula, Quercus, and Sorbus) were encoded as levels of a categorical variable used to predict $\mathbb{E}[Y]$. The original trapezoidal fuzzy numbers were transformed into Beta-type fuzzy numbers using a procedure designed to minimize information loss from the original fuzzy sets, as described in \cite{calcagni2022modeling}.

Given the nature of the observed fuzzy data, we hypothesized two candidate models for the non-fuzzy component, one involving the Beta distribution $\mathcal{M}_1: f_Y(y; \boldsymbol{\theta}_y) = \mathcal Be(y;\mu_i\phi,\phi-\phi\mu_i)$ and another defined using the LogitNormal distribution $\mathcal{M}_2: f_Y(y; \boldsymbol{\theta}_y) = \mathcal{LGN}(y;\mu_i,\sigma)$. These distributions were selected for their suitability in modeling bounded rating data. Convergence was achieved for all chains and both modeling cases, as indicated by $\widehat{R}$ values equal to 1. The effective sample sizes (ESS) further supported the robustness of the results. Specifically, the bulk-ESS values ranged from 7474 to 7945, ensuring reliable estimates of central tendencies, while the tail-ESS values, ranging from 9089 to 9415, confirmed sufficient exploration of the posterior distribution's tails. Finally, the model $\mathcal{M}_1$ was selected based on its lower WAIC value ($\text{WAIC}_{\mathcal{M}1} = -9184$, $\text{WAIC}_{\mathcal{M}_2} = -7631$). The PPC steps (i)–(iii) were applied $B = 500$ times for the chosen model. 

\begin{figure}[htbp]
\centering
	%\resizebox{15cm}{!}{\input{fig4app1.tex}}
	\includegraphics[scale=0.8]{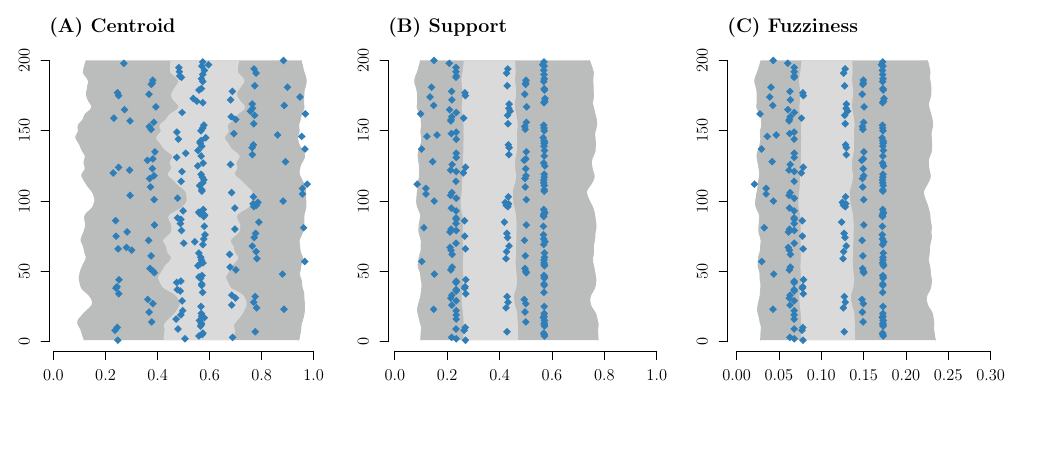}
	\caption{Reforestation data analysis -- Panels (A)-(C) illustrate the interquartile range (light gray area), the max-min range (dark gray area), and the defuzzified observed data (blue dots) for each of the posterior predictive statistics. The horizontal axis represents the scale of the response variable $Y$, while the vertical axis corresponds to the statistical units}
	\label{fig4app1}
\end{figure}

Figure \ref{fig4app1} presents the results for the three key statistics. Overall, the posterior predictive analysis suggests a strong goodness-of-fit, with the predicted fuzzy data closely matching the observed data across all evaluated metrics (centroids: $\text{CP} = 0.93$, $\text{bP} = 0.02$; supports: $\text{CP} = 0.92$, $\text{bP} = 0.02$ and fuzziness: $\text{CP} = 0.92$, $\text{bP} = 0.03$). These results indicate that the final model effectively captures the key characteristics of the observed data with a high degree of accuracy.

%%%%%%%%%%%%

\subsection{Cognitive response times (Kim \& Bishu, 1998)}\label{sec6.2}

The second dataset comes from a study on the cognitive response times of nuclear power plant control room crews during abnormal events \cite{kim1998evaluation}. Various performance factors, including stress levels and the potential for misdiagnosis, influence these response times. As noted by Kim and Bishu \cite{kim1998evaluation}, cognitive times can be effectively represented as fuzzy numbers, capturing the subjective nature of human responses under stress. For this study, the entire sample of $n=75$ units was utilized, along with three continuous predictors to model $\mathbb{E}[Y]$: {Inside control room experience} ($X_1$), {Outside control room experience} ($X_1$), and {Education} ($X_3$). The original triangular fuzzy numbers were transformed into Beta-type ones using the procedure detailed in \cite{calcagni2022modeling}.

\begin{figure}[htbp]
\centering
	%\resizebox{15cm}{!}{\input{fig5app1.tex}}
	\includegraphics[scale=0.8]{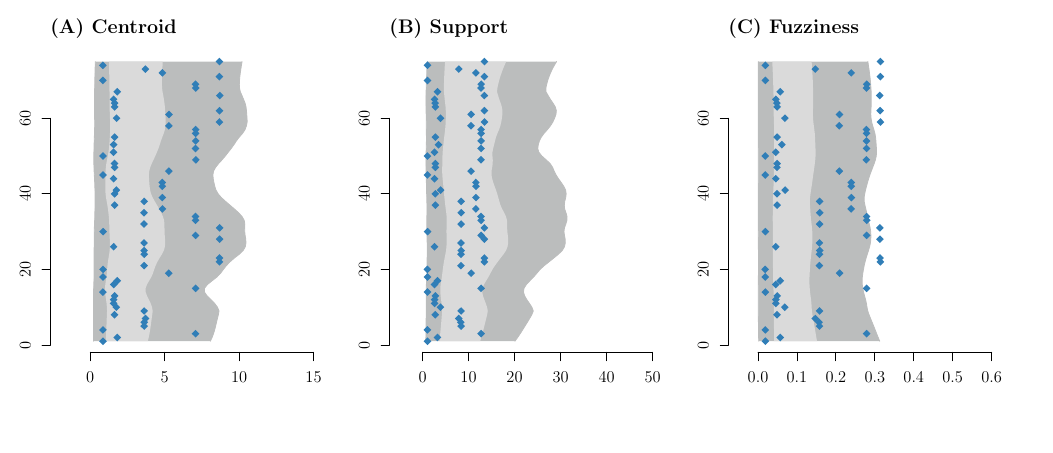}
	\caption{Cognitive response times analysis -- Panels (A)-(C) illustrate the interquartile range (light gray area), the max-min range (dark gray area), and the defuzzified observed data (blue dots) for each of the posterior predictive statistics. The horizontal axis represents the scale of the response variable $Y$, while the vertical axis corresponds to the statistical units}
	\label{fig5app1}
\end{figure}

Considering the nature of the response variable, we followed the Kim \& Bishu's choice to adopt the {Lognormal} distribution for modeling $f_Y(y;\boldsymbol{\theta}_y) = \mathcal{LN}(y;\mu_i,\sigma^2)$. In this case, as $Y\in\mathbb R^+$, bounds for the modes $m_1,\ldots,m_n$ were computed by using 0-cuts of the observed fuzzy numbers (see Section \ref{sec3.3}). All the chains simulated here converged, with the bulk-ESS ranging from 7772 to 8000 and the tail-ESS from 9113 and 9503. The PPC steps (i)–(iii) were applied $B = 500$ times for the chosen model. Figure \ref{fig5app1} presents the results for the three statistics under consideration. The posterior predictive analysis suggests a marginally adequate goodness-of-fit for the estimated model (centroids: $\text{CP} = 0.99$, $\text{bP} = 0.03$; supports: $\text{CP} = 0.80$, $\text{bP} = 0.18$; fuzziness: $\text{CP} = 0.90$, $\text{bP} = 0.10$).

\subsection{Employee performance assessments (Gong et al., 2018)}\label{sec6.3}

The third dataset, used for external validation of our approach, pertains to employee performance assessments \cite{Gong2018395}. It comprises $n = 30$ triangular fuzzy observations related to performance evaluation and includes a set of fuzzy triangular predictors, such as work quality and inability to endure job stress. As in previous applications, the triangular fuzzy numbers were transformed into Beta-type fuzzy numbers using the procedure outlined in \cite{calcagni2022modeling}. For this analysis, only work quality was considered as a predictor of $\mathbb{E}[Y]$.

Since $Y$ represents a bounded rating variable, the Kumaraswamy distribution $f_Y(y; \boldsymbol{\theta}_y) = \mathcal K(y;\alpha_i,\nu)$ was chosen to model the non-fuzzy component, with the following parametrization $\nu = (1+\exp(-\phi))^{-1}$ and $\alpha_i = -\log(1-0.5^\nu)/\log(1+\exp(-\mathbf X\boldsymbol\beta))$. Convergence was achieved for all chains, with bulk-ESS ranging from 7011 to 8101 and tail-ESS from 8897 to 9302. The PPC steps (i)--(iii) were applied $B = 500$ times for the chosen model. Figure \ref{fig6app1} presents the results for the three evaluated statistics. Overall, the posterior predictive analysis suggests a good fit for the estimated model, with the predictive metrics achieving a satisfactory level of accuracy (centroids: $\text{CP} = 0.99$, $\text{bP} = 0.04$; supports: $\text{CP} = 0.90$, $\text{bP} = 0.03$; fuzziness: $\text{CP} = 0.90$, $\text{bP} = 0.02$).

\begin{figure}[htbp]
\centering
	%\resizebox{15cm}{!}{\input{fig6app1.tex}}
	\includegraphics[scale=0.8]{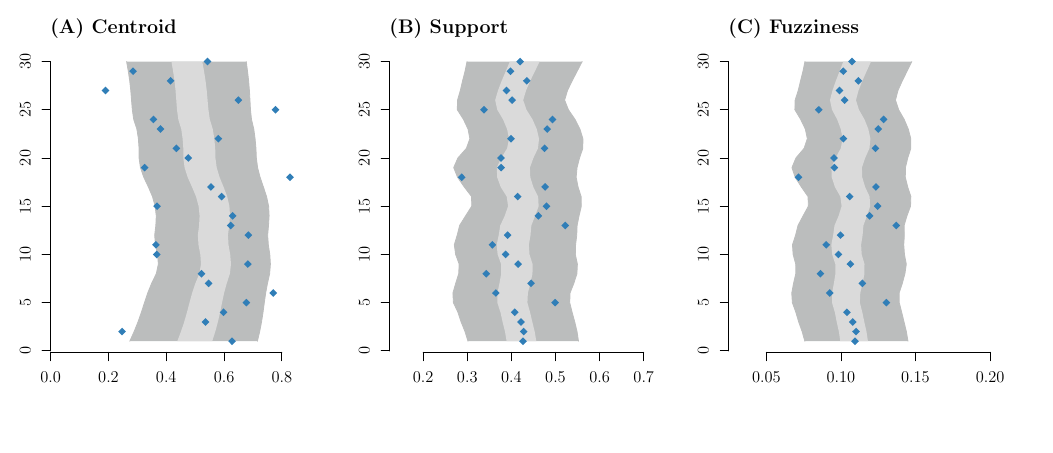}
	\caption{Employee performance assessments -- Panels (A)-(C) illustrate the interquartile range (light gray area), the max-min range (dark gray area), and the defuzzified observed data (blue dots) for each of the posterior predictive statistics. The horizontal axis represents the scale of the response variable $Y$, while the vertical axis corresponds to the statistical units}
	\label{fig6app1}
\end{figure}

\subsection{Case study}\label{sec6.4}

Sexual intimacy is crucial to human relationships, contributing significantly to emotional and social well-being. Its importance has been particularly evident during the COVID-19 pandemic, which brought unprecedented challenges to relationships through increased stress, isolation, and disrupted routines \cite{pennanen2021sexual}. These circumstances underscored the need to understand the predictors of sexual intimacy better, especially as many couples faced heightened uncertainty in their dynamics. Assessing the determinants of sexual intimacy is inherently complex, as self-reported responses are often influenced by biases such as decision uncertainty and social desirability \cite{hook2003close}. This case study aims to address these challenges by examining how relational and behavioral factors, such as perceived partner responsiveness and sexual desire, contribute to sexual intimacy. The original data refers to an extensive survey conducted in Flanders in 2021 that investigated four main factors related to sexual intimacy \cite{van2018associations}. As this application has an illustrative purpose only, we drew a smaller sample of participants and focused on the following variables only: age (in years, \texttt{age}), relationship duration (in years, \texttt{rel\_length}), desire (composite score, \texttt{sex\_desire}), gender of the partner (dichotomous, \texttt{gender\_partner}), partner responsiveness (composite score, \texttt{respo\_partner}), and perceived intimacy (self-reported, \texttt{intimacy}). The sample used here consisted of $n=318$ participants, including 232 women (mean age: 34.15; sd: 11.95) and 95 men (mean age: 31.07; sd: 9.58). The mean relationship duration was 7.74 years (sd: 8.91 years) for women and 7.71 years (sd = 8.33 years) for men. As the perceived intimacy was the variable to be modeled here, the original ratings collected on a Likert-type scale were fuzzified according to the fuzzy-IRTree methodology \cite{calcagni2022psychometric}, which carried out a $(1,5)$-Beta-type fuzzy variable. 

The approximated Gibbs sampler described in Section \ref{sec4} was used on the final dataset, with five independent chains of 4000 samples each (burn-in period: 2000 samples). As for the previous applications, weakly informative priors were used. Since the response variable consists of rating data, the $(1,5)$-Beta distribution with the mean-precision parametrization was used to model the non-fuzzy component, namely $f_Y(y;\boldsymbol\theta_y) = \mathcal{B}e(y;\mu\phi,\phi-\mu\phi)_{[1,5]}$. The variables \texttt{age}, \texttt{rel\_length}, \texttt{sex\_desire}, \texttt{gender\_partner}, and \texttt{respo\_partner} were used in the linear predictor $\eta_i$ for $\mu_i = \Exp{Y_i} = g^{-1}(\eta_i)$ (as usual, the logistic function was used as invere link function $g^{-1}$). All the chains reached the convergence according to the Gelman and Rubin's $\widehat R$ index and were subsequently used in the data analysis. Table \ref{tab3app} reports some of the posterior statistics, the 95\% Highest Posterior Density Interval (HPDI) bounds, and the Effective Sample Size (ESS) diagnostics. Figure \ref{fig6app1} shows the marginal posterior densities for the model parameters.

\begin{table}[htbp]
\centering
\caption{Case study -- Posterior quantiles and 95\% HPDIs for the model parameters. Here, $\beta_0$ stands for the intercept of $\mu$ and codifies the baseline level \texttt{gender\_partner = male}. The parameter $\phi$ is the precision parameter of the non-fuzzy Beta model.} 
\label{tab3app}
\resizebox{12cm}{!}{
\begin{tabular}{lcccccc}
  \hline
 & mean & sd & HPDI lb & HPDI ub & ESS-bulk & ESS-tail \\ 
  \hline
$\beta_0$ & 1.15 & 0.03 & 1.09 & 1.22 & 7552.72 & 9265.22 \\ 
  $\beta_\texttt{age}$ & 0.04 & 0.04 & $-0.04$ & 0.13 & 8581.15 & 9306.59 \\ 
  $\beta_\texttt{rel\_length}$ & $-0.03$ & 0.04 & $-0.11$ & 0.07 & 8529.86 & 9484.83 \\ 
 $\beta_\texttt{sex\_desire}$ & 0.04 & 0.03 & $-0.03$ & 0.10 & 8264.60 & 9089.53 \\ 
  $\beta_\texttt{gender\_partner:female}$ & 0.40 & 0.03 & 0.33 & 0.46 & 7792.56 & 9127.77 \\ 
  $\beta_\texttt{respo\_partner}$ & $-0.01$ & 0.03 & $-0.07$ & 0.06 & 8655.80 & 9599.46 \\ 
  $\phi$ & 21.34 & 0.11 & 2.85 & 3.27 & 7016.59 & 8151.62 \\ 
   \hline
\end{tabular} }
\end{table}

The posterior results indicate that \(\widehat{\beta}_0 = 1.15\) with a standard deviation of \(\sigma_{\widehat{\beta}_0} = 0.03\), and the 95\% HPDI is \([1.09, 1.22]\) (the intercept codifies the case \(\texttt{gender\_partner} = \texttt{male}\)). For \(\texttt{age}\), the estimate is \(\widehat{\beta}_\texttt{age} = 0.04\) with \(\sigma_{\widehat{\beta}_\texttt{age}} = 0.04\), and the HPDI is \([-0.04, 0.13]\), indicating no strong evidence for an effect. Similarly, \(\texttt{rel\_length}\) has an estimate of \(\widehat{\beta}_\texttt{rel\_length} = -0.03\) with \(\sigma_{\widehat{\beta}_\texttt{rel\_length}} = 0.04\), and the HPDI is \([-0.11, 0.07]\), also suggesting no strong evidence for an effect. For \(\texttt{sex\_desire}\), the estimated value is \(\widehat{\beta}_\texttt{sex\_desire} = 0.04\) with \(\sigma_{\widehat{\beta}_\texttt{sex\_desire}} = 0.03\), and the HPDI is \([-0.03, 0.10]\), indicating a weak effect that is not strongly supported by the data. Conversely, \(\texttt{gender\_partner:female}\) shows a clear positive effect with \(\widehat{\beta}_\texttt{gender\_partner:female} = 0.40\), \(\sigma_{\widehat{\beta}_\texttt{gender\_partner:female}} = 0.03\), and an HPDI of \([0.33, 0.46]\), indicating a robust and substantial effect. Finally, \(\texttt{respo\_partner}\) has an estimated value of \(\widehat{\beta}_\texttt{respo\_partner} = -0.01\) with \(\sigma_{\widehat{\beta}_\texttt{respo\_partner}} = 0.03\), and the HPDI \([-0.07, 0.06]\), suggesting no evidence of a significant effect. The precision parameter is estimated as \(\phi = 21.34\), with a standard deviation of \(\sigma_\phi = 0.11\) and an HPDI \([2.85, 3.27]\). The posterior results suggest that sexual intimacy is positively associated with having a female partner, as indicated by the significant positive effect of the partner's gender. Additionally, sex desire and age exhibit weak effects, with their 95\% HPDIs narrowly including zero, suggesting potential but inconclusive associations. Other predictors, such as the relationship length and the perceived partner responsiveness, show negligible effects on the outcome.

\begin{figure}[htbp]
\centering
	%\resizebox{10cm}{!}{\input{fig7app1.tex}}
	\includegraphics[scale=0.8]{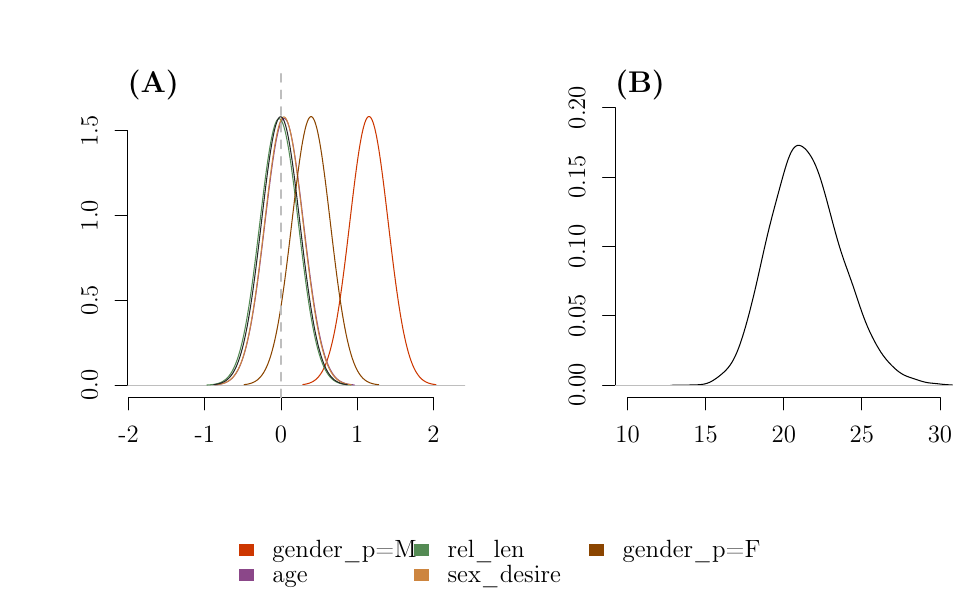}
	\caption{Case study -- Marginal posterior densities for the model parameters: (A) regression coefficients on the linear predictor scale, (B) precision coefficient}
	\label{fig6app1}
\end{figure}

To further investigate whether the relationship between perceived intimacy and partner responsiveness varies across the gender of partners (as one might expect in many circumstances \cite{birnbaum2016intimately}), the linear predictor \(\eta\) of the non-fuzzy component was extended to include the interaction \texttt{gender\_partner} \(\times\) \texttt{respo\_partner}. The new model \(\mathcal{M}_2\) was then estimated and compared with the previous model \(\mathcal{M}_1\) via the WAIC measure. Although the model including the interaction performed slightly better (\(\text{WAIC}_{\mathcal{M}_1} = 4399.381\) and \(\text{WAIC}_{\mathcal{M}_2} = 4374.075\)), the interaction effect was negligible (\(\widehat{\beta}_\texttt{interaction} = 0.022\), \(\sigma_{\widehat{\beta}_\texttt{interaction}} = 0.16\), \(\text{HPDI} = [-0.11, 0.07]\)). Overall, the results suggest that while gender differences play a significant role in predicting sexual intimacy, the hypothesized moderating effect of partner responsiveness across genders does not have meaningful support in the data.

The Supplementary Materials (Sections 2.1-2.2) contain extended numerical and graphical results for this case study, including a comprehensive evaluation of the results obtained using the proposed conditional model compared to those from a selection of state-of-the-art regression methods for fuzzy data.

\section{Conclusions}\label{sec7}

This paper introduces a conditional probabilistic framework for the statistical analysis of bounded fuzzy data, providing a consistent and comprehensive approach to handling fuzzy variables. While numerous fuzzy statistical tools exist, they often lack a unified framework integrating general, widely applicable parametric probability models for fuzzy random variables. This gap has hindered their practical application, particularly in delivering broadly applicable inferential results. Building on the premise that fuzzy data arise from a process akin to statistical coarsening, we propose a regression-based approach that links observed fuzzy statistics -- such as modes and spreads -- to their underlying statistical model counterparts. This model also accommodates covariates within a more general approximated Gibbs sampling-based solution, enhancing its flexibility and applicability. Through simulations and case studies, we illustrate the effectiveness of the proposed method, showcasing its ability to incorporate fuzziness within a structured inferential setting. Furthermore, leveraging a Bayesian approach enhances both the interpretability and inferential capacity of fuzzy data analysis.

Significantly, the conditional approach proposed in this work is grounded in an epistemic interpretation of fuzziness, in which observed fuzzy values are treated as imprecise representations of latent, unobserved random variables. This assumption does not rely on any specific cognitive mechanism on the part of the observer; instead, it provides a statistically principled way to model non-stochastic imprecision. Consequently, the proposed framework is applicable across multiple domains, such as sensory evaluations, questionnaire-based ratings, and decision support systems, where individuals often provide fuzzy assessments due to limited knowledge or intrinsic measurement imprecision. In all these cases, the blurring mechanism serves to approximate how non-fuzzy random quantities are transformed into fuzzy responses through the measurement process.

An additional contribution of this paper pertains to the ongoing debate on the simulation of fuzzy numbers. While traditional methods have predominantly relied on deterministic or arithmetic approaches for generating fuzzy numbers (see \cite{grzegorzewski_amcs2020,pgmr2022,PGMR2024AMS} for a detailed discussion), our approach adopts a probabilistic framework grounded in Bayesian methods. This integration embeds the simulation process within a broader inferential paradigm, offering a unified statistical representation that seamlessly handles data simulation and analysis. In particular, our method aligns naturally with the parametric regression framework, enhancing its applicability in modeling and inference. Other epistemic sampling processes considered previously in the literature \cite{Grzegorzewski2022,grzegorzewski_amcs2020,Colubi2002,Lubiano2017} are usually more straightforward in their nature, albeit without the significant ``Bayes flavor'' existing in the proposed approach \eqref{eq1}--\eqref{eq3}.

By and large, this work emphasizes that, as in other areas of mathematical and statistical modeling, a solid foundation for the proposed constructions remains essential. In particular, we think that any data analysis with fuzzy numbers need to be properly formulated within a probabilistic framework, which ensures that the methodology developed for analyzing imprecise data based on random fuzzy numbers is statistically well-justified. While many of the solutions available in the literature on fuzzy data analysis are largely heuristic, the approaches proposed in this work are firmly grounded in the methodology of standard statistical inference. Special attention is given to the beta-type fuzzy numbers considered here, which, in our opinion, deserve particular emphasis due to their ease of use, interpretation, generation, understanding, and visualization, as well as their convenience for Bayesian analysis considering that most of the existing literature restricts itself to trapezoidal fuzzy numbers.

Despite its strengths, our approach is not without limitations. Notably, it reduces fuzziness to a collection of summary statistics -- albeit meaningful ones -- potentially sacrificing some of its full expressive power. Additionally, the proposed method does not incorporate an integrated possibilistic framework for statistical inference, which could better capture fuzzy data's inherent richness and complexity. However, it is essential to recognize that the scientific community currently lacks a fully possibilistic approach that matches the expressiveness and inferential capabilities of probabilistically grounded statistical frameworks. This trade-off, though common in statistical modeling, parallels challenges in other fields, such as network data analysis, where simplifications are often made to enable inference. This raises a fundamental question: To what extent does replacing fuzzy numbers with summary statistics represent real progress, and could this shift be more meaningfully interpreted within the broader statistical paradigm? After six decades of research in fuzzy set theory, it is worth reconsidering the role of fuzzy statistics in modeling imprecise data and whether its goals can be effectively incorporated into existing statistical approaches (e.g., Bayesian Networks). The challenge of addressing and ultimately resolving this question falls to future generations of scholars.

\vspace{2cm}
%\clearpage
\section*{Appendix}

\noindent Equation \eqref{eq3a} is derived by directly applying the law of iterated expectations to the random variable  $M|S,Y$
\[
	\Exp{M} = \Exp{\Exp{M|Y,S}} = \Exp{\frac{SY}{SY-SY+S}} = \Exp{Y}.
\]
Similarly, one obtains \eqref{eq3b} by the law of iterated variances
\begin{align*}
	\Var{M} &= \Var{\Exp{M|Y,S}} + \Exp{\Var{M|S,Y}} \\
			&= \Var{\frac{SY}{SY-SY+S}} + \Exp{\frac{(1-Y)Y}{S+1}} \\
			&= \Var{Y} + \Exp{\frac{1}{S+1}}\Exp{Y(1-Y)} \\
			&= \Var{Y} + \Exp{S^*}(\Exp{Y}-\Exp{Y^2}) \\			
			&= \Var{Y} + c(\Exp{Y}-\Var{Y}-\Exp{Y}^2),
\end{align*}
where the term $\Exp{Y^2}$ has been rewritten using the identity $\Exp{Y^2} = \Var{Y} + \Exp{Y}^2$. The scaling factor $c$ represents the expectation of the random variable $S^*$, which is defined by the following density function
\begin{equation*}
	f_{S^*}(x;\alpha,\beta) = \frac{\beta^{-\alpha} e^{\frac{x-1}{x\beta}} \left(-1 + \frac{1}{x}\right)^{\alpha-1}}{x^2 \Gamma(\alpha)},
\end{equation*}
where $x\in(0,1)$, $\alpha>0$ and $\beta>0$. The expectation does not have a closed-form solution, except in the simplest case where $\alpha<1$, for which it is given by $\Exp{S^*} = \beta^{-\alpha} e^{\frac{1}{\beta}}\Gamma\left( 1-\alpha,\frac{1}{\beta} \right)$. For all other cases, numerical integration must be employed.

\clearpage
\bibliographystyle{plain}
\bibliography{biblio}

\end{document}